\documentclass[preprints,article,accept,moreauthors,pdftex]{Definitions/mdpi} 
\DeclareUnicodeCharacter{3BC}{$\mu$}

\firstpage{1} 
\pubvolume{1}
\issuenum{1}
\articlenumber{0} 
\pubyear{2022}
\copyrightyear{2022}
\datereceived{} 
\dateaccepted{} 
\datepublished{} 
\hreflink{https://doi.org/} % If needed use \linebreak
\doinum{10.3390/electronics11244201}
\pdfoutput=1

\Title{seL4 Microkernel for virtualization use-cases: Potential directions towards a standard VMM}
\TitleCitation{seL4 Microkernel for virtualization use-cases: Potential directions towards a standard VMM}

\Author{Everton de Matos $^{1,*}$\orcidA{}, Markku Ahvenjärvi$^{1,2}$}

\AuthorNames{Everton de Matos and Markku Ahvenjärvi}

\AuthorCitation{de Matos, E.; Ahvenjärvi, M.}
\address{% 
$^{1}$ \quad Technology Innovation Institute - Secure Systems Research Center, Abu Dhabi, United Arab Emirates;\\
$^{2}$ \quad Unikie Oy, Tampere, Finland
}

\corres{Correspondence: everton.dematos@tii.ae}

\abstract{%A single paragraph of about 200 words maximum. For research articles, abstracts should give a pertinent overview of the work. We strongly encourage authors to use the following style of structured abstracts, but without headings: (1) Background: place the question addressed in a broad context and highlight the purpose of the study; (2) Methods: describe briefly the main methods or treatments applied; (3) Results: summarize the article's main findings; (4) Conclusions: indicate the main conclusions or interpretations. The abstract should be an objective representation of the article, it must not contain results which are not presented and substantiated in the main text and should not exaggerate the main conclusions.
Virtualization plays an essential role in providing security to computational systems by isolating execution environments. Many software solutions, called hypervisors, have been proposed to provide virtualization capabilities. However, only a few were designed for being deployed at the edge of the network, in devices with fewer computation resources when compared with servers in the Cloud. Among the few lightweight software that can play the hypervisor role, seL4 stands out by providing a small Trusted Computing Base and formally verified components, enhancing its security. Despite today being more than a decade with seL4 microkernel technology, its existing userland and tools are still scarce and not very mature. Over the last few years, the main effort has been put into increasing the maturity of the kernel itself and not the tools and applications that can be hosted on top. Therefore, it currently lacks proper support for a full-featured userland Virtual Machine Monitor, and the existing one is quite fragmented. This article discusses the potential directions to a standard VMM by presenting our view of design principles and feature set needed. This article does not intend to define a standard VMM, we intend to instigate this discussion through the seL4 community.}


\begin{document}

%%%%%%%%%%%%%%%%%%%%%%%%%%%%%%%%%%%%%%%%%%
\section{Introduction}
\label{sec:introduction}
An in-depth study by Transparency Market Research has found that embedded systems in IoT have witnessed incredible progress. The study has projected the embedded system market to advance at a CAGR (Compound Annual Growth Rate) of 7.7\% from 2022 to 2031 \cite{BLOOM22}. Moreover, the growing trend of virtualization for embedded systems in the IT sector is a major force for the expansion of avenues in the embedded system market. Virtualization stands out as an approach to providing portability and security to computational systems. It has become popular as a solution for high-powered machines, as servers in the Cloud Computing paradigm. However, in recent years, it has been popularizing the use of virtualization at the Edge of the network in embedded devices \cite{GER08}. Virtualization makes it possible to have different and isolated virtual machines (i.e., execution environments) in the same platform, providing security by separation, in which one environment does not have access to the resources of a neighbor environment. The hypervisor is the piece of software that creates and runs the virtual machines in a system \cite{BAU15}. There is a variety of flavors for hypervisors available nowadays. 

The Protected KVM \cite{DEA20} and the Dom0-less Xen \cite{STA19} are two recent open-source initiatives that have been trying to improve the security of mainstream hypervisors like KVM (Kernel-based Virtual Machine)\footnote{https://www.linux-kvm.org/page/Main\_Page} and Xen\footnote{https://xenproject.org/}, respectively. However, they are sacrificing the performance or the feature set for which those hypervisors were originally designed. Microkernel-based designs like seL4 rapidly became revolutionary since they bring security from the ground-up, from its conception to its implementation. seL4 is an operating system microkernel, but it is also a hypervisor with sufficient mechanisms to be possible to build anything atop \cite{HEI09}.

The seL4 is a small and simple microkernel-based type-1 hypervisor. It follows the principle of least privilege with capability-based access control mechanisms, performant inter-process communication channels, and user-space services \cite{HEI20}. seL4 works in two different modes although sharing the same concepts, either as an OS or as a Hypervisor, depending on a set of design-time configurations. The microkernel standalone requires user-space services and Virtual Machine Monitors (VMMs) that run along with guest OSes. 

Using seL4 in complex production environments \cite{SERV22} brings new challenges often quite different from research-oriented environments. There is a fragmentation problem since different companies are using their own closed-source tools and user-space libraries. This problem could likely be solved by creating a VMM implementation standard that focuses more on the existing business goals of seL4 ecosystem members. The effort would benefit from the input and authorship of seL4 integrators \cite{MEM22}, which are active commercially focused teams in the seL4 ecosystem, including but not limited to Cog Systems\footnote{https://cog.systems/}, Technology Innovation Institute\footnote{https://www.tii.ae/secure-systems}, Hensoldt Cyber\footnote{https://hensoldt-cyber.com/}, DornerWorks\footnote{https://dornerworks.com/}, NIO\footnote{https://www.nio.com/}, and Ghost\footnote{https://www.ghostautonomy.com/}. Each of those businesses has different goals, but there is still a common baseline and philosophy binding them around the VMM when using seL4 as a hypervisor. 

At a high level, there are gaps in the seL4 stack, specifically the VMM, userspace, and tooling, which complicate matters for integrators attempting to meet real-world customer use cases. Not all business opportunities require a solution using a VM (Virtual Machine) architecture, but those that do quickly become complex and would benefit enormously from an established standard or reference baseline. The lack of a robust and consistent VMM for seL4 has created a highly fractured environment. Most integrators have their own specialized customer use cases, and they have found that the quickest path is to use a forked and modified VMM. This practice may have short-term benefits to that integrator. Still, it does not allow the community to benefit from commonality and guarantees that the fork will quickly get old and out of sync with the mainline. For instance, there will be VMM fork features that overlap and which should be implemented in a standard way for the sake of ongoing community benefit and maintenance. 

As a community, a VMM standard can be created which is consistent, maintainable, and works for all the varied business focuses. Critically, this effort must get the conceptual buy-in of the seL4 Foundation and the larger seL4 community for it to be successful going forward. Additionally, a reference implementation, properly maintained, would help to solidify and promote the concept and to provide consistency to complex projects using seL4.

Recent publications discuss the challenges and characteristics of several hypervisors in different application domains \cite{CIN22} \cite{WUL21} \cite{AAL21}. Also, the use and deployment of seL4-based solutions have been the subject of study in a large number of scientific papers in the last few years \cite{GER20} \cite{VAND18} \cite{VAN19} \cite{MILL20} \cite{SUD22}. However, to the best of our knowledge, there is no article in the literature discussing the challenges toward a standard VMM in seL4. 

The main objective of this article is to help in overcoming the seL4 VMM fragmentation problem. In light of this, the present article has the following goals:
\begin{itemize}
    \item To gather the community in building a standard VMM as a shared effort by highlighting the discussion about the challenges and potential directions of a standard VMM;
    \item This article does not intend to define a standard VMM. It intends to present the potential key design principles and feature set support toward seL4 VMM standardization. The items shown in this paper can be the basis for an extended version with a more comprehensive list of required properties and features.
\end{itemize}

The contributions of this article can be summarized as (i) to present a discussion on the seL4 microkernel being used as a hypervisor, and its comparison with other traditional hypervisors, (ii), to present potential directions for seL4 VMM development and adoption by the community, and (iii) to discuss the next steps on seL4 virtualization apart from the VMM development, as the use of an API to connect with different VMMs and the next steps of formal verification in this spectrum.

The rest of this article is presented as follows. Section \ref{sec:why} introduces seL4 and presents its characteristics. Section \ref{sec:background} presents the background definitions surrounding the virtualization and seL4 topics. Section \ref{sec:standardVMM} presents the Design Principals and Feature Support Tenets towards a standardized seL4 VMM. Section \ref{sec:discussion} presents discussion topics that should be considered towards a seL4 standard VMM. Finally, Section \ref{sec:conc} concludes the article.

%%%%%%%%%%%%%%%%%%%%%%%%%%%%%%%%%%%%%%%%%%
\section{Why seL4?}
\label{sec:why}

seL4 is a member of the L4 family of microkernels that goes back to the mid-1990s \cite{HEI20} \cite{ELP13}. It uses capabilities, which allows fine-grained access controls and strong isolation guarantees. The Trusted Computing Base, or TCB, with seL4 is small with 9-18k SLOC (source lines of code), depending on CPU architecture, and it was the first general purpose OS to be formally verified. seL4 also features very fast IPC (Inter-Process Communication) performance - something that is very important for microkernels. According to seL4 FAQ \cite{FAQ22}, it is the fastest microkernel in a cross-address-space message-passing (IPC) operation.

Many of the hypervisors have as their main strengths other aspects than security. This impacts the architecture (e.g. monolithic) and design decisions. In this regard seL4 with it’s fine grained access model and strong isolation guarantees outperform others. The formal verification further adds proof and  credibility and makes it even more unique. Thus, seL4 has a solid security model and story, backed by formal verification.

The seL4 is a general-purpose microkernel with proven real-time capabilities that provides system architectural flexibility. The security and safety critical components can be run natively in the user space of seL4 hypervisor. This also applies to the components with real-time requirements. The strong spatial and temporal isolation guarantees that the system components - and untrusted VMs - are unable to interfere with each other.

seL4 is used and being developed by a growing number of companies and hobbyists, with only a few hypervisors, as KVM and Xen, outperforming seL4 in this regard. Most of the Open Source Hypervisors (OSS) have a small engaged community, and/or the development solely depend on the interest of a single individual or company. Community is one of the most important aspects for a successful open source operating system, and hypervisor. Moreover, there are hypervisors being developed by a single company. In this case, the development takes place in private repositories, and only the selected features are published as snapshots to public repositories. The dominance of a single company makes these projects unattractive for other companies. This, for example, hinders the development of architecture as well as hardware support in general. In seL4 environment, the seL4 foundation\footnote{https://sel4.systems/Foundation/home.pml}\cite{FOUN22} ensures neutrality, and all seL4 development takes place on public repositories.

%%%%%%%%%%%%%%%%%%%%%%%%%%%%%%%%%%%%%%%%%%
\section{Background}
\label{sec:background}
This Section presents various concepts that are used during the article. It starts from the definition and comparison between a microkernel and a monolithic kernel. It discusses the definition and disambiguation of the terms Virtual Machine and Virtual Machine Monitor under the umbrella of virtualization. It discusses different hypervisors approaches and its VMMs. Finally, it discusses the VMM definition under seL4 environment and a variety of its components on the virtualization side, as for instance CAmkES\footnote{https://docs.sel4.systems/projects/camkes/} (Component Architecture for Microkernel-based Embedded Systems) and Core Platform\footnote{https://trustworthy.systems/projects/TS/sel4cp/}.

\subsection{Microkernel \& Monolithic kernel}
The kernel is the indispensable and therefore most important part of an operating system \cite{ROCH04}. An operating system will consist in two main parts: (i) kernel space, and (ii) user space. The kernel space runs in a higher privilege level than the user space. Any code executing in privileged mode can bypass security, and is therefore inherently part of a system’s trusted computing base (TCB) \cite{RUSH81}. 

There are two different concepts of kernels: monolithic kernel and microkernel. The monolithic kernel runs a every basic system service in kernel space level of privilege. Examples of those basic system services are process and memory management, interrupt handling and I/O communication, and file system \cite{ROCH04} \cite{NID19}. The kernel size, lack of extensibility, and poor maintainability are the three main disadvantages of placing all fundamental services in the kernel space. The microkernel was created with the idea of reduce the kernel to basic process communication and I/O control, and let the other system services reside in user space in form of normal processes \cite{ROCH04}. This approach reduces the TCB of the kernel itself, thus reducing the attack surface \cite{BIG18}. 

Although monolithic kernels tend to be more generic and easy to use than microkernels, its large TCB increases the vulnerabilities at the kernel space. As an example, there are about 22.7M SLOC for the whole Linux kernel but 16.4M SLOC (71.9\%) of them are device drivers \cite{HAO22}. On the other hand, as an example, seL4 microkernel has around 9-18k SLOC \cite{FAQ22}. It is easier to ensure the correctness of a small kernel, than a big one. That way, stability issues are simpler to solve with that approach \cite{ROCH04}. Moreover, It has been argued that the microkernel design, with its ability to reduce TCB size, contain faults and encapsulate untrusted components, is, in terms of security, superior to monolithic systems \cite{BIG18} \cite{HOH04}.

\subsection{Virtualization}
Virtualization is a technique that allows several operating systems to run side-by-side on given hardware \cite{CHI05} \cite{POP74}. Virtualization brings different kinds of benefits to the environment that it is deployed. One of the benefits would be the heterogeneity that it can bring, being possible to deploy various operating systems and applications in the same hardware \cite{TIB21}. Moreover, it improves the system's security by achieving security by separation \cite{MORA20} \cite{MART17}. It is achieved as each operating system has its own space, not having an explicit connection with others, keeping software instances isolated. Nevertheless, virtualization requires a software layer responsible for system management, known as a hypervisor.

The hypervisor is a software layer responsible for managing the hardware and explicitly making it available to the upper layers \cite{SMI05}. It has privileged access to the hardware resources and can allocate it accordingly to the operating systems. Examples of hardware resources or devices are: storage memory, network device, I/O devices, etc. For security reasons, the hardware should not be shared directly by different operating systems. However, the hypervisor can provide virtual copies of the same hardware to other operating systems \cite{RUSS08}. Many computer architectures have specific privilege levels to run the hypervisor, such as EL2 on ARM and HS-mode on RISC-V. Examples of hypervisors are Xen, KVM, ACRN\footnote{https://projectacrn.org/}, Bao\footnote{http://www.bao-project.org/}, and seL4. 

The hypervisors can be categorized into type-1 and type-2. The type-1 hypervisors runs on bare metal (i.e., directly on the host machine's physical hardware) and type-2 hypervisors, also called hosted hypervisors, runs on top of an operating system \cite{NAK19}. The type-1 hypervisors are considered more secure by not relying on a host operating system. KVM is an example of type-2 hypervisor by running on Linux kernel while seL4 is an example of type-1 hypervisor. 

\begin{figure}[hbt]
    \centering
    \includegraphics[scale=0.8]{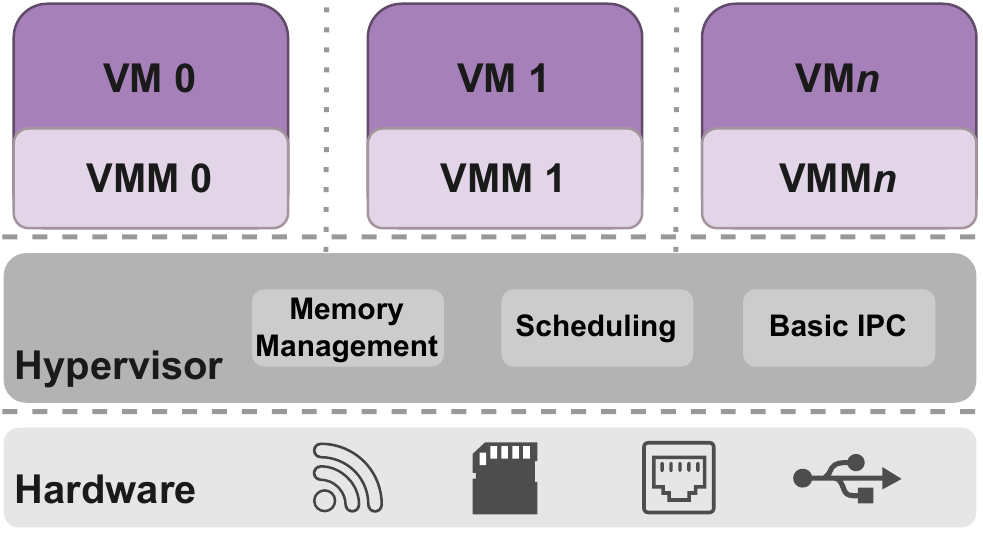}
    \caption{Overview of a virtualization environment considering seL4 type-1 hypervisor}
    \label{fig:simple_virt}
\end{figure}

The Figure \ref{fig:simple_virt} presents a high-level overview of the components present in a virtualization environment considering seL4 hypervisor: hardware resources or devices, hypervisor, Virtual Machine Monitor, and Virtual Machine. The seL4 microkernel when used as a type-1 hypervisor, provides only basic functionality (i.e., memory management, scheduling tasks, basic IPC), pushing more complex functionalities, as device drivers, to the upper layers. Type-2 hypervisors will have a larger code base with more complex functionalities embedded into it \cite{BIG18}. 

%\begin{adjustwidth}{-\extralength}{0cm}

The Virtual Machine Monitor (VMM) is a piece of software that interacts with the hypervisor in the virtualization environment. It has its own responsibilities apart from the hypervisor. The VMM is a user space program that provides emulation for virtual devices and control mechanisms to manage VM Guests (virtual machines) \cite{AZM11}. The VMM enables the virtualization layer to create, manage, and govern operating systems \cite{ROS05}. By running at the user space, the VMM runs at privilege level EL0 on ARM and U-mode on RISC-V. Examples of VMMs are Firecracker\footnote{https://firecracker-microvm.github.io/}, \textit{crosvm}\footnote{https://chromium.googlesource.com/chromiumos/platform/crosvm/}, QEMU\footnote{https://www.qemu.org/}. Depending on the hypervisor and on the characteristics of the deployed environment, it is possible to have one or multiple VMMs. A common approach is to have one VMM per each operating system Virtual Machine. 

Each operating system sits inside a Virtual Machine (VM). A VM behaves like an actual operating system from the point of view of the user, being possible to run applications and interact with it \cite{TIC10}. From the point of view of the hypervisor, a VM has access to a specific set of hardware resources managed by the hypervisor. It is the VMM that makes the bridge from the hardware resources of the hypervisor to make them available to the VM by managing the backend operations \cite{XU14}. From the scalability perspective, it is possible to have multiple VMs in a virtualization environment, where each VM is isolated from the other by principle. The quantity of VMs depends on the amount of physical resources available for such an environment.

\subsection{Related hypervisors and VMMs}
Apart from seL4, there are other open source hypervisors available in the market. KVM and Xen are examples of traditional hypervisors that have been in the market for more than 15 years and were deployed in different solutions \cite{BAR03} \cite{CHI10}. While both hypervisors are widely used, feature rich and well supported, the huge TCB makes them vulnerable.

KVM is a type-2 hypervisor that added virtualization capabilities to Linux. KVM is integrated in the Linux kernel, thus benefiting from reusing many Linux functionalities such as memory management and CPU scheduling. The downside of it is the huge TCB that comes along KVM. The KVM was originally built for x86 architecture and then ported to ARM \cite{DALL14}. The KVM on ARM implementation has been split in the so-called Highvisor and Lowvisor. The Highvisor lies in ARMs kernel space (EL1) and handles most of the hypervisor functionalities. The Lowvisor resides in hypervisor mode (EL2) and is responsible for enforcing isolation, handling hypervisor traps and performing the world switches (context execution switches between VMs and host) \cite{RAH15}.

Xen is defined as a type-1 hypervisor. The x86 version of Xen, is a bare-metal hypervisor that supports both fully virtualized and para-virtualized guests. On ARM, the code for Xen is reduced to one type of guest which uses para-virtualized drivers and the ARM virtualization extensions \cite{RAH15}. The Xen hypervisor resides in hypervisor mode. On top of it, everything is executed as a guest placed in different domains. The most privileged domain is called Dom0, it has access to hardware and runs Linux to manage other guests, named DomU\footnote{https://wiki.xenproject.org/wiki/DomU}. A DomU is the counterpart to Dom0; it is an unprivileged domain with (by default) no access to the hardware. DomU’s use Dom0’s para-virtualized services through Xen PV calls. Recently, the Dom0-less variant was introduced. With Dom0-less\footnote{https://xenproject.org/2019/12/16/true-static-partitioning-with-xen-dom0-less/}, Xen boots selected VMs in parallel on different physical CPU cores directly from the hypervisor at boot time. Xen Dom0-less is a natural fit for static partitioning, where a user splits the platform into multiple isolated domains and runs different operating systems on each domain. 

Traditionally, KVM and Xen hypervisors were designed to be deployed at the Cloud Computing level to provide virtualization to high-density machines. However, recent solutions were developed to use those kinds of hypervisors also at the Edge level \cite{YAS21} \cite{RAM16}, being able to have such virtualization solutions in devices with less processing power than the servers at the Cloud \cite{HW08} \cite{STA12} \cite{DALL14}. There are also hypervisors that were designed in a lightweight manner, with the intention to be applied in resource-constrained environments at the Edge level. Examples of lightweight hypervisors are Bao \cite{MAR20} and ACRN \cite{LI19}, among others. 

Bao is a lightweight bare-metal hypervisor designed for mixed-criticality systems. It strongly focuses on isolation for fault-containment and real-time behavior. Its implementation comprises a thin-layer of privileged software leveraging ISA virtualization support to implement the a static partitioning hypervisor architecture \cite{MAR20}. ACRN targets itself to IoT and Egde systems, placing a lot of emphasis to performance, real-time capabilities and functional safety. ACRN currently only supports x86 architectures, and as it is mainly backed by Intel, support to other architectures may not appear any time soon \cite{LI19}. 

Gunyah\footnote{https://github.com/quic/gunyah-hypervisor} is a relatively new hypervisor by Qualcomm. It is a microkernel design with capability access controls. Gunyah being a new project has a very limited HW support, and practically non-existent community outside Qualcomm. KVMs\footnote{https://github.com/jkrh/kvms} is an \textit{aarch64} specific hypervisor, building upon popular KVM, bringing a lot of flexibility for example in terms of choice of VMMs. Thanks to a small size, it is possible to formally verify hypervisor EL2 functionality \cite{WEI21}. While there are a lot of benefits, it is limited to on CPU architecture, and maintaining KVMs patch series across several versions of Linux kernel may become an issue.

KVM relies in user space tools such as the Quick Emulator (QEMU) \cite{BELL05} to serve as VMM and instantiating virtual machines. In the KVM paradigm guests are seen by the host as normal POSIX processes, with QEMU residing in the host userspace and utilizing KVM to take advantage of the hardware virtualization extensions \cite{RAH15}. Other VMM can be use on top of KVM, as Firecracker, Cloud Hypervisor and crosvm. Firecracker uses the KVM to create and manage microVMs. Firecracker has a minimalist design. It excludes unnecessary devices and guest functionality to reduce the memory footprint and attack surface area of each microVM \cite{AGA20}. Cloud Hypervisor focuses on exclusively running modern, cloud workloads, on top of a limited set of hardware architectures and platforms. Cloud workloads refers to those that are usually run by customers inside a cloud provider. Cloud Hypervisor is implemented in Rust and is based on the \textit{rust-vmm}\footnote{https://github.com/rust-vmm} crates. The crosvm VMM is intended to run Linux guests, originally as a security boundary for running native applications on the Chrome OS platform. Compared to QEMU, crosvm does not emulate architectures or real hardware, instead concentrating on para-virtualized devices, such as the VirtIO \cite{VIRT18} standard.

\subsection{seL4 VMM}
An Operating System (OS) microkernel is a minimal core of an OS, reducing the code executing at higher privilege to a minimum. The seL4 is a microkernel and hypervisor capable of providing virtualization support \cite{HEI20}. It has a small trusted computing base (TCB), making a minor surface attack compared to traditional hypervisors such as KVM and Xen.

The seL4 supports virtualization by providing specifically two libraries: (i) \textit{libsel4vm}, and (ii) \textit{libsel4vmmplatsupport} \cite{VIR22}. The first (i) is a guest hardware virtualization library for x86 (ia32) and ARM (ARMv7/w virtualization extensions \& ARMv8) architectures. The second (ii) is a library containing various VMM utilities and drivers that can be used to construct a guest VM on a supported platform. These libraries can be utilized to construct VMM servers through providing useful interfaces to create VM instances, manage guest physical address spaces and provide virtual device support (e.g., VirtIO Net, VirtIO PCI, VirtIO Console). Projects exist that make use of the seL4 virtualization infrastructure, supporting the provision of virtualization environments. Examples of those kinds of projects are CAmkES and Core Platform.

The CAmkES project is a framework for running virtualized Linux guests on seL4 for ARM and x86 platforms. The \textit{camkes-vm} implements a virtual machine monitor (VMM) server, facilitating the initialization, booting and run-time management of a guest OS \cite{CAMVMM22}. The CAmkES project provides an easy way to run different virtualization examples with one or more VMs and different applications. It also provides a way how to passthrough devices in such environments. One drawback of such a framework is that it is only possible to run static VMs, in which the VM configuration should be defined at design time.

When using CAmkES, a system is modelled as a collection of interconnected software components, as CAmkES follows a component-based software engineering approach to software architecture. These software components are designed with explicit relationships between them and provide interfaces for explicit interaction \cite{CAM22}. The development framework provides: (i) a domain-specific language (DSL) to describe component interfaces, components, and whole component-based systems, (ii) a tool that processes these descriptions to combine programmer-provided component code with generated scaffolding and glue code to build a complete, bootable, system image, and (iii) full integration in the seL4 environment and build system.

CAmkES proved to be too complex, static and maintenance intensive. Because of this reason, many projects and companies have rolled their own user space. As the VMM is in the user space, the challenges and limitations are imminent in the virtualization too. To remedy the situation, the seL4 community is introducing seL4 Core Platform \cite{COR22} \cite{COR22b}, or seL4cp, and seL4 Device Driver Framework\footnote{https://sel4.atlassian.net/browse/RFC-12}, or sDDF. The two new components are attempts to fix the shortcomings of CAmkES. This also means that the VMM parts will be changed significantly too.

The Core Platform provides the following abstractions: protection domain (PD), communication channel (CC), memory region (MR), and notification and protected procedure call (PPC). A VM is a special case of a PD with extra, virtualization-related attributes. The whole virtual machine appears to other PDs as just a single PD, i.e. its internal processes are not directly visible \cite{COR22b}. A PD runs an seL4CP program, which is an ELF (Executable and Linkable Format) file containing code and data, both of which are exposed as memory regions and mapped into the PD. The original version of the seL4CP was fully static, in that all code had to be fixed at system build time, and PDs could not be restarted. The addition of dynamic features is in progress \cite{COR22c}. The seL4 Device Driver Framework (sDDF) provides libraries, interfaces and protocols for writing/porting device drivers to run as performant user level programs on seL4. The sDDF also aims to be extended to a device virtualization framework (sDVF) for sharing devices between virtual machines and native components on seL4.

Even though the seL4 VMM exists and is available to use, it lacks in providing essential features for virtualization support in complex scenarios. Moreover, its fragmentation by different closed-source deployments makes the mainline depreciate fast. Thus, it is necessary to discuss the desired features for such a standard VMM.

%%%%%%%%%%%%%%%%%%%%%%%%%%%%%%%%%%%%%%%%%%
\section{Philosophy of a Standard VMM}
\label{sec:standardVMM}
It should be immediately obvious that even a community as small as the commercial users of seL4 will have difficulty agreeing to an all-encompassing standard. Thus, what is proposed is to establish a driving philosophy for the design of a baseline VMM rather than prescribe a specific system architecture. There is the need to discuss the possible missing features of the existing seL4 VMM \cite{VIR22} concerning a standard VMM, more so than a prescription for the right way to do it. Indeed, this will entail recommending high-level architecture patterns but cannot lock an adopter into specific implementations. Each adopting integrator will inevitably start from the new standard and refine the implementation for their use case. One size does not fit all, so customization will always occur. The effort here is to close the gap between the current VMM baseline and the point of necessary deviation. Refinement should only be necessary to cover specific requirements and edge cases highly unlikely to appear in multiple projects across the integrator community. 

For this discussion, driving philosophical concepts can be roughly binned into Design Principles and Feature Support Tenets. The Design Principles and Feature Support Tenets were defined based on features present in already available VMMs (see Section \ref{sec:background}) and the technical challenges they posed. The Design Principles were also defined based on the most common Quality Attributes for Embedded Systems \cite{SHER08} \cite{CLE03} \cite{OLI13} \cite{BIA15}. Moreover, the Feature Support Tenets are also based on the open challenges on seL4 virtualization domain, accordingly to open discussions in seL4 community channels\footnote{https://sel4.discourse.group/}$^{,}$\footnote{https://lists.sel4.systems/postorius/lists/devel.sel4.systems/}. A deeper discussion about the Design Principles and Feature Support Tenets will be needed before implementations at seL4 mainline. This list intends to be a starting point for discussing such topics.

\subsection{Design Principles}
Five major design principles are recommended as potential directions towards the standard VMM. They are motivated to be open, modular, portable, scalable, and secure.

\subsubsection{Official and Open Source}
The existing seL4 VMM \cite{VIR22} employs an open-source license, and any new implementations under the proposed standard should remain in accordance with this approach. This applies to all the code up to the point of necessary differentiation. Individual integrators should always retain the ability to keep closed-sourced their highly specialized or trade secret modifications. This strikes a balance between business needs such as maintaining a competitive edge and fully participating in a collaborative community around a common baseline. Open sourcing the standard VMM is essential for the seL4 community to engage collaboratively and improve the VMM by either contributing to the source code repository or using and learning from it.

It is recommended to place the standard VMM baseline under the purview of the seL4 Foundation to benefit from the structure and governance of that organization. The desire is that it will gain in stature as well, as the current VMM is a second-class citizen in the community. Alongside the source code, the Foundation should periodically publish reports about major updates and possible new directions as new technologies mature. In this way, it will help to maintain a long-term roadmap to incorporate new features such as ARMv9 Realms \cite{XUP22}, for instance.

\subsubsection{Modular: Maintainable and Upgradable}
It is expected that the standardized VMM would be deployed in heterogeneous environments and scenarios under quite varied use cases. This will require flexibility in aspects such as system architecture, hardware requirements, performance, etc. It is essential to follow a modular design approach to guarantee the applicability of the VMM in any of those variants. 

In implementing the VMM modularly, it is essential to achieve its readability by following the C4 (Context, Containers, Components, and Code) model, for instance. The C4 model is an "abstraction-first" approach to diagramming software architecture \cite{VAZ20}. It decomposes the system so community members can pick and choose components for their project. The Context shows how the software system in scope fits into the environment around it. Containers inside a Context define the high-level technical building blocks. A Component is a zoom-in to an individual Container and shows its responsibilities and implementation details. Finally, Code is a specific description of how a Container is implemented. The modular approach makes it possible for integrators to define the Context, Containers, Components, and Code that must be pieced together for a VMM to support specific features, making their VMM highly customized to their end goal.

\subsubsection{Portable: Hardware independence}
With the vast number of supported platforms by the seL4 kernel \cite{SUP22}, the VMM should also be generic enough to support them. It can be a lofty goal, but the standard VMM should be designed and written with hardware abstraction layers such that it can compile for a minimum of the ARM, x86, and RISC-V Instruction Set Architectures. In this way, the standard VMM is not explicitly linked to a specific set of hardware characteristics. Of course, different ISAs may impose architectural differences. However, there is the need for a minimal and modular VMM that could be easily moved from 4 core ARM SoC (big.LITTLE) to a 48-core Thread Ripper AMD x86, as an example. The standard VMM could be seen as a baseline for different hardware implementations. Obviously, the baseline will not take advantage of all the platforms' hardware features. However, it can be used for Proof-of-Concept implementation and learning purposes for being easy to deploy on different platforms.

Additionally, it is essential to consider and accommodate the rather large differences architecture-wise, even with the same ISA implementation. For example, Codasip\footnote{https://codasip.com/} and SiFive\footnote{https://www.sifive.com/} implementations of RISC-V have non-ignorable differences, while ARM implementations from Qualcomm, Samsung, and NXP exhibit wildly different behavior \cite{PIN19}. Though SoC vendors may be compliant with the ISA specification, there usually is some collection of deviations or enhancements present, often implemented as a black-box binary. Areas of concern include control of the system’s Memory Management Units, Generic Interrupt Controller, TPM/TEE, secure boot process, and access to the appropriate privilege level for the seL4 kernel (e.g., EL2 for Qualcomm-ARM).

\subsubsection{Scalable: Application-agnostic}
A standard VMM should be scalable in the sense that it needs to be able to support several applications running on top for different specific purposes. Different applications may have a distinct set of requirements such as performance, safety, security, or real-time. The VMM should be able to meet those requirements and provide a way for the applications to reach them. Moreover, the VMM should guarantee that the applications will run as expected, being able to initiate and finish the tasks successfully. A VMM scheduler should be responsible for balancing the loads and ensuring that no application (i.e., thread) is left unattended.

The scalability of the systems is also tied to their performance. In light of this, it is essential that the VMM supports from one to an arbitrary number of processing units or cores. The existing seL4 VMM does not support multiprocessing and consequently highly restricts the number of applications that can be run atop. Enabling multiprocessing would help achieve better performance, thus improving the scalability of the system performance as a whole. We discuss in detail the possibilities to enable multicore VMM further in this paper in the Multicore \& Time Isolation section. 

\subsubsection{Secured by Design}
\begin{figure}[hbt]
    \centering
    \includegraphics[scale=0.7]{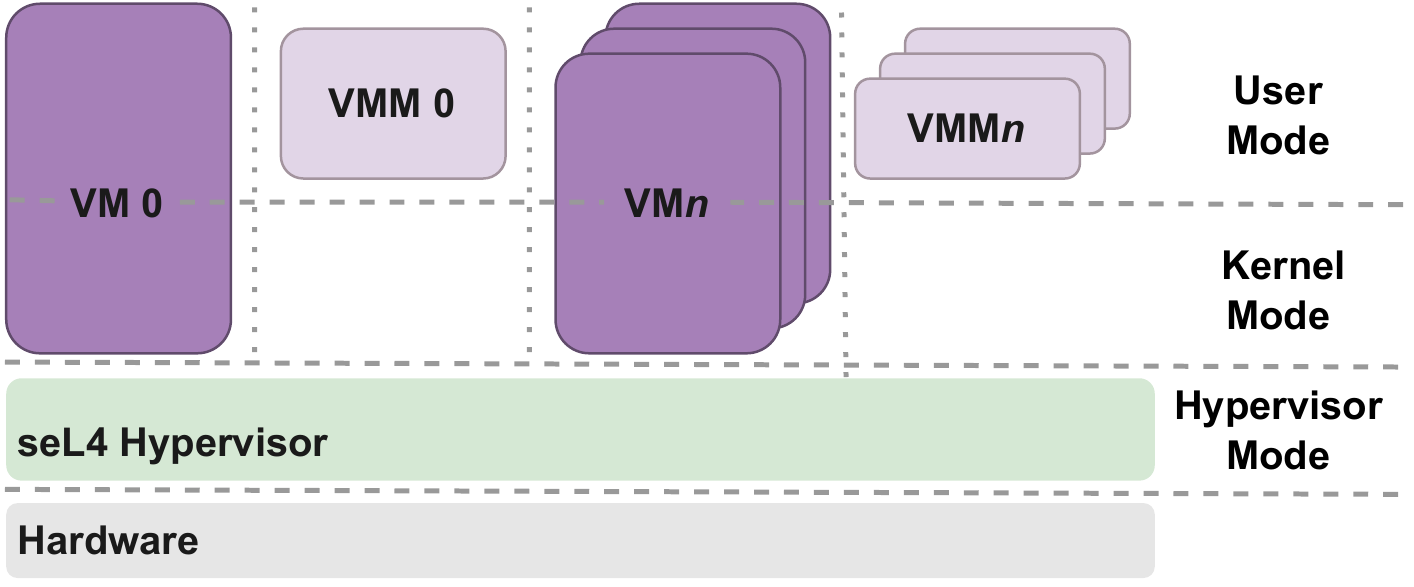}
    \caption{Example of an architecture with one VMM per VM}
    \label{fig:whitepaper_1vmm1vm}
\end{figure}

A standard seL4 VMM implementation should support one VMM instance per VM. Even though this approach is well followed by most of the integrators and supported by seL4, it is important to highlight its benefits. This approach improves both scalability and security of the solution. If a guest OS is compromised, it opens an attack vector toward the VMM. However, the risk is limited if there is a dedicated VMM per VM. The other VMs, their VMMs, and guest OSes are completely isolated by the stage 2 translation. This assumes a formally verified kernel and that the translation tables or the memory areas the tables point to are distinct for each VM. Though this approach is already common today, some integrators do not always implement it for time-to-market pressure, reusable code, or other unusual circumstances. Support for this design should be standardized so that the enabling code can be considered boilerplate and easily consumed. Figure \ref{fig:whitepaper_1vmm1vm} shows a representation of a secured by design architecture, with one VMM per VM. Even though the VMM has more direct interaction with the hypervisor, it is placed in the User Mode. The VMs are present at both User and Kernel modes, as they can have applications and drivers, respectively.

\subsection{Feature Set Support Tenets}
Four major features are recommended as potential directions towards the standard VMM to support hardware mechanisms and provide security and performance benefits.

\subsubsection{System Configuration} 
Currently, there are two main approaches to facilitate the system configuration when running virtual environments on top of seL4. The first to be introduced was CAmkES, that stands for Component Architecture for microkernel-based Embedded Systems \cite{KUZ07}. The second one is seL4 Core Platform (seL4CP) \cite{COR22}. The Core Platform, which was recently introduced, intends to be the standard for such virtual environments on top of seL4. Thus, the CAmkES is being deprecated.

CAmkES is a software development and runtime framework for quickly and reliably building microkernel-based multiserver (operating) systems \cite{KUZ07}. Currently, using the CAmkES framework with the VMM will result in a fully static configuration. The VMs must be defined and configured during the build. This also includes defining the RAM area. It is designed to achieve security guarantees so as not to allow post-build modifications to the number of running VMs and their interconnections. This is a highly desirable aspect when the use case calls for it. However, it can be inflexible and even short-sighted when the nature of the user experience requires dynamic configuration, i.e. no dynamic start/stop/restart capability.

It is often necessary to have a more dynamic seL4-based environment for the purpose of allowing better usability, modularity, or even scalability. The Core Platform is an operating system (OS) personality for the seL4 microkernel. The Core Platform makes seL4-based systems easy to develop and deploy within the target areas. It can be used to bring up VMs on top of seL4. Core Platform promises to deliver dynamic features to the seL4 environment \cite{COR22}. However, it is still in progress with ongoing virtualization features in development\footnote{https://github.com/Ivan-Velickovic/sel4cp/tree/virtualisation\_support}. The Trustworthy Systems - UNSW\footnote{https://trustworthy.systems/about/} group also intends to formally verify two core aspects of the seL4 Core Platform\footnote{https://trustworthy.systems/projects/TS/sel4cp/verification}: (i) correctness of the implementation, i.e. its abstractions function as specified, and (ii) correctness of the system initialisation, i.e. the collection of underlying seL4 objects are fairly represented by the system specification.

A new VMM standard should enhance the existing static build approach with a build-time specification stating that dynamic configurations are also permitted. They could be limited by providing build-time parameters for acceptable configurations. To achieve a dynamic environment, it should be possible to use the seL4 mechanisms for transferring/revoking capabilities to the entities during runtime, providing a potential implementation mechanism for this feature. It may also be an option to build a core common component to serve as an “admin VM” for dynamic configurations, even subjecting it to some degree of formal methods verification. This is anticipated to be an area of much research and prototyping to achieve the desired balance of security and flexibility. 

\subsubsection{Multicore \& Time Isolation}
One of the key aspects of virtualization is the need for efficiency, where multiprocessing configurations play an important role. Although multicore support is a complex engineering task, it should be supported in its simplest shape to avoid contention and potential deadlocks. Different physical CPUs (pCPUs) can be enabled by the kernel (in a Symmetric Multiprocessing - SMP configuration) in order to allocate them to a different system running threads according to the use-case application requirements. Next, we present potential multi-core configurations that a standard VMM should be able to support using a clear multiprocessing protocol:

\begin{itemize}
    \item Direct Mapping Configuration: multiple single-core VMs running concurrently and physically distributed over dedicated CPUs. Figure \ref{fig:amp} shows the representation of the Direct Mapping Configuration approach.
    
    \begin{figure}[hbt]
    \centering
    \includegraphics[scale=0.77]{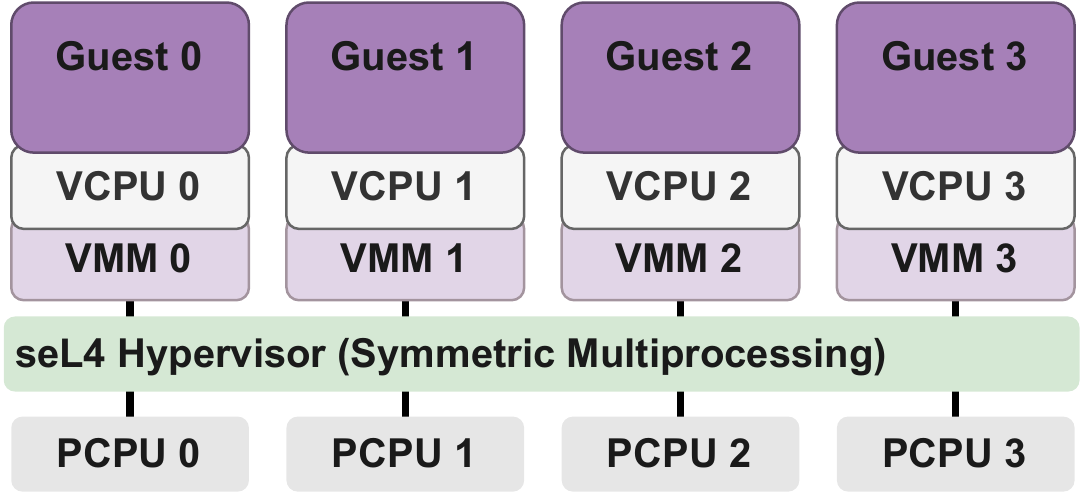}
    \caption{Direct Mapping Configuration overview}
    \label{fig:amp}
    \end{figure}

    \item Hybrid Multiprocessing Configuration: it can have multiple single-core VMs running in dedicated CPUs as the Direct Mapping Configuration, however, it can also have multicore VMs running in different CPUs. Figure \ref{fig:hmp} shows the representation of the Hybrid Multiprocessing Configuration approach.
    
    \begin{figure}[hbt]
    \centering
    \includegraphics[scale=0.77]{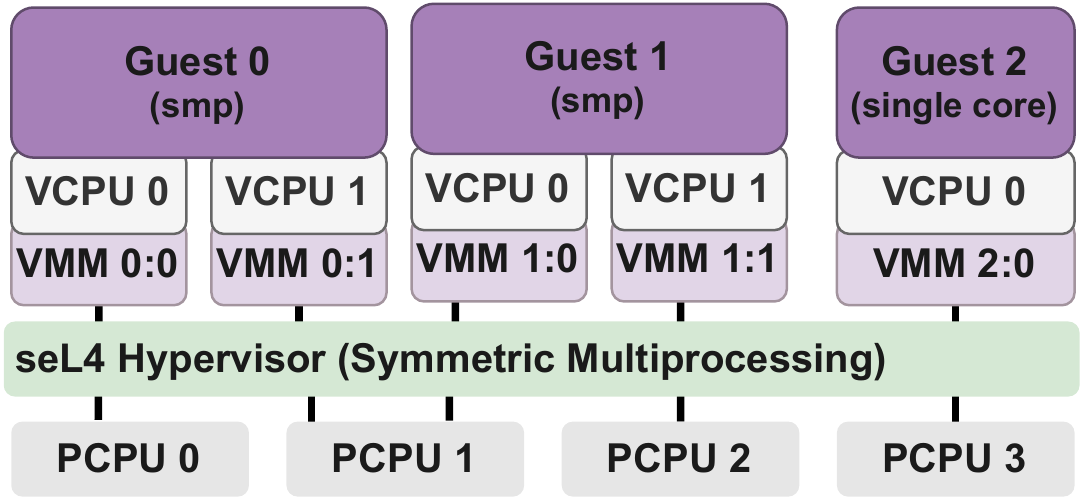}
    \caption{Hybrid Multiprocessing Configuration overview}
    \label{fig:hmp}
    \end{figure}
    
\end{itemize}

The two depicted configurations are examples for future reference of a standard VMM, but it is not strictly limited. Most, if not all, current and near-future use cases are covered by a model where there are multicore VMMs that are pinned to exclusive cores and unicore VMMs that can be multiplexed on a core. Ideally, it would be up to the system designer to decide which configuration to use. It could be either static or dynamic, enabling switching from a given configuration to another in run-time. The selected configuration will affect several threads in execution. In the seL4 context, threads can be running either Native apps, OSes, and/or VMMs. The former is typically used to run device drivers or support libraries. OSes are using threads running over virtual abstractions, or VMs, while VMMs are creating and multiplexing these abstractions to be able to encapsulate OSes. They all require an abstraction representing the pCPU time but differ from the supported execution level and their scope over other system components. For example, a VMM can access the VM internals but not the opposite. Other features that are likely required by multicore design are:

\begin{itemize}
    \item \textbf{vCPUs Scheduling:} The ability to schedule threads on each pCPU based on their priority, credit-based time slicing, or budgeting depending on the algorithm selected. As an example, It could be a design configuration whether it supports vCPU migration (a vCPU switching from pCPU id:0 to id:1) with also the possibility to tie up a set of the vCPUs to pCPUs. Another potential configuration is the static partitioning one, where all the vCPUs are assigned to pCPUs at design-time and are immutable at run-time. In addition, having dynamic and static VMs configuration in a hybrid mode could be something to support. A multiprocessing protocol with acquire/release ownership of vCPUs should be supported. The seL4 kernel has a scheduler that chooses the next thread to run on a specific processing core, and is a priority-based round-robin scheduler. The scheduler picks threads that are runnable: that is, resumed, and not blocked on any IPC operation. The scheduler picks the highest-priority, runnable thread (0~255). When multiple TCBs are runnable and have the same priority, they are scheduled in a first-in, first-out round-robin fashion. The seL4 kernel scheduler could be extended for the VMMs.
    \item \textbf{pIRQ/vIRQs ownership:} physical interrupts (pIRQs) shall be virtualized (vIRQs) and require a multiprocessing protocol with simple acquire/release ownership of interrupts per pCPU/vCPU targets. Besides, support hardware-assisted interrupt-controllers with multicore support is required.
    \item \textbf{Inter-vCPU/inter-pCPU communication:} Another key aspect of multiprocessing architectures is the ability to communicate between pCPUs. Also, with equal importance, communication between vCPUs results in not only inter-pCPU but also inter-vCPU communication. Communication is very important in multiprocessing protocols, but it should be designed in a way that is simple to verify and validate.
\end{itemize}

\subsubsection{Memory Isolation}
As stated before, the secure by design principle is strongly based on memory isolation. Memory isolation is critical to enforce the security properties such as VMs confidentiality and integrity. Hardware-enforced and partial microkernel access-controlled memory translation and protection between VMs/VMMs and Native Apps are key security requirements for security-critical use-cases. Support for hardware-assisted virtualization (extended Page Tables or second-stage) MMU should be an integral part of the standard VMM. Next, are some features for future reference that can leverage such hardware for memory isolation: (i) configurable VM Virtual Address Space (VAS); (ii) device memory isolation; and (iii) cache isolation.

\begin{itemize}
    \item \textbf{Configurable VM Virtual Address Space:} Multiple virtual Address Spaces are an important feature supported by high-end processors and have the same paramount importance for hardware-assisted virtualization. There should be different Virtual Address Spaces for different software entities: Hypervisor, VMM, and their respective VMs. User-controlled and configurable address spaces are important features for VMs. For example, (i) setting up a contiguous virtual address space ranges from fragmented physical memory as well as small memory segments shared with other VMs, (ii) hiding physical platform memory segments or devices from the VMs, (iii) no need to recompile a non-relocatable VM image.
    
    \item \textbf{Device memory isolation by hardware-support or purely software:} Devices that are connected to the System-on-Chip (SoC) bus interconnection and are masters can trigger read and write DMA transactions from and to the main memory. This memory, typically DRAM, is physically shared and logically partitioned among different VMs by the hypervisor. Some requirements could be met in a standard VMM: (i) a device can only access the memory of the VM it belongs to; (ii) the device could understand the virtual AS of its VM; and (iii) the virtualization layer could intercept all accesses to the device and decode only those that intend to configure its DMA engine in order to do the corresponding translation if needed, and control access to specific physical memory regions. In order to meet these three requirements a standard VMM requires support for either an IOMMU (with one or two stage translation regimes) or software mechanisms for mediation.
    
    \item \textbf{Cache isolation through page-coloring:} Micro-architectural hardware features like pipelines, branch predictors, and caches are typically available and essential for well performant CPUs. These hardware enhancements are mostly seen as software-transparent but currently leaving traces behind and opening up backdoors that can be exploited by attackers to break memory isolation and consequently compromising the memory confidentiality of a given VM. One mitigation for this problem is to apply page coloring in software and could be an optional feature supported by a standard VMM.  Page coloring is meant to map frame pages to different VMs without colliding into the same allocated cache line. A given cache allocated by a VM cannot evict a previously allocated cache line by another VM. This technique, by partitioning the cache in different colors, can protect to some extent (shared caches) against timing cache-based side channel attacks, however, it strongly depends on some architectural/platform parameter limitations such as cache size, number of ways and page size granularity used to configure the virtual address space. L1 cache is typically small and private to the pCPU while L2 cache is typically bigger and seen as the last level of cache that is shared among several pCPUs. It would be possible to assign a color to a set of VMs based on their criticality level. For example, assuming the hardware limits the system to encode up to 4 colors, where one color can be shared by a set of non-critical VMs, other for real-time VM for deterministic behavior, and the other two for a security- and performance-critical VM that requires increased cache utilization and at the same isolation against side-channel attacks.
\end{itemize}

\subsubsection{Hypervisor-agnostic I/O Virtualization and its derivations}
Many security use-cases require virtualization environments with reduced privilege such that only specific VMs, called driver VMs, can directly access hardware resources while the others, called User VMs, run in a driverless mode since device drivers are seen today as a major source of bugs. A compromise caused by exploitation of a driver bug can be contained in its own VM. Typically, in such environments, any VM that will potentially run unknown code and/or untrusted applications may require isolation from key device drivers sequestered into their dedicated VMs. Inter-VM communication, including access to the devices, must be done by proxy over well-known and managed interfaces. This approach requires a combination of VM kernel modifications and VMM modules to be able to communicate and share basic hardware devices over virtual interfaces. 

The OASIS collaboration community manages the set of VirtIO standards \cite{VIRT18} that are implemented to various degrees by Linux and Android. Given the excellent support, it is recommended to adopt VirtIO implementations for multiple interfaces in the standard VMM. Support for standardized VirtIO server implementations in the VMM would be a meaningful complement to guest OS clients. For instance, the VirtIO-Net server in the VMM could store a table of MAC addresses, creating a virtual switch. In the case of the VirtIO-Block server, the VMM could terminate VirtIO-Block requests so that address mappings are not known by the user-facing guest OS, then start up another request to the VM containing the device driver to perform the actual write. For instance, in complex architectures with more than one guest OS accessible from the user perspective, VMM VirtIO servers could also handle multiplexing access to various devices between VMs, creating a “multi-persona” capability.

Among the possibilities of implementing VirtIO interfaces, the following items present examples of how it can be used and integrated with a standard VMM: 

\begin{itemize}
    \item VirtIO can be used for interfacing VMs with host device drivers. It can support VirtIO driver \textit{backends} and \textit{frontends} on top of seL4. VirtIO interfaces can be connected to open-source technologies such as QEMU, \textit{crosvm}, and \textit{Firecracker}, among others. In this scenario, the open-source technologies will execute in the user space of a VM different from the one using the device itself. This approach helps in achieving reusability, portability, and scalability. Figure \ref{fig:virtio_whitepaper} shows the representation of such an approach considering a VirtIO Net scenario in which a Guest VM consumes the services provided by a back-end Host VM.
    
    \begin{figure}[hbt]
    \centering
    \includegraphics[scale=0.77]{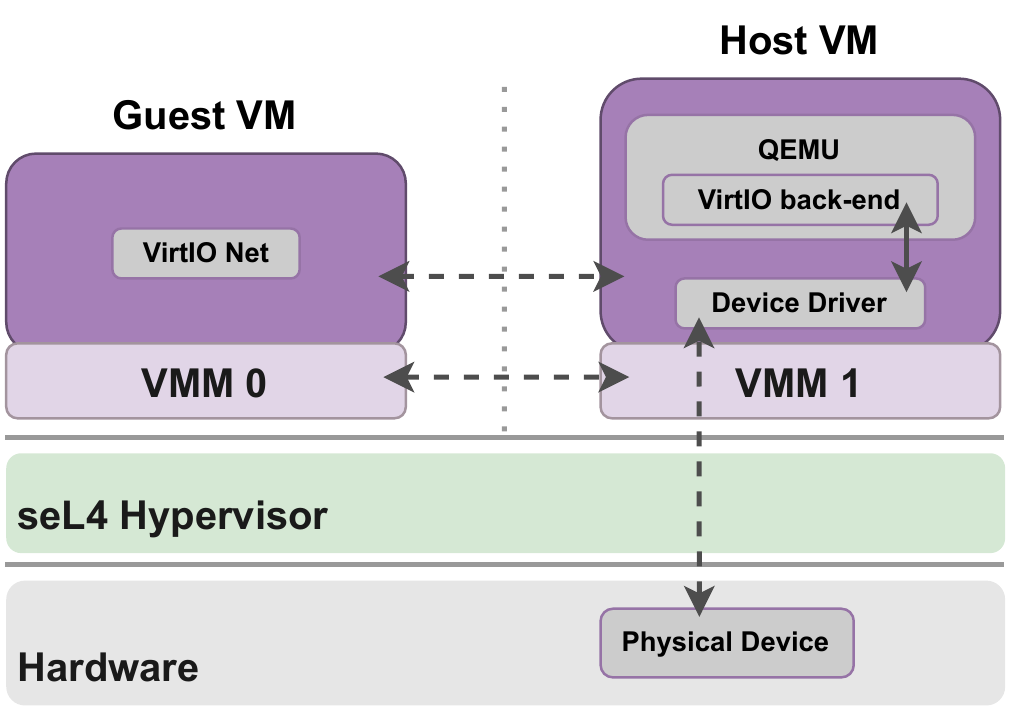}
    \caption{VirtIO drivers example on top of seL4 hypervisor}
    \label{fig:virtio_whitepaper}
    \end{figure}

    \item VirtIO interfaces can be connected to formal verified native device drivers. The use of such kinds of device drivers increases the security of the whole system. Moreover, the verified device drivers can be multiplexed to different accesses, switching device access between multiple clients. The multiplexer is transparent to native clients, as it uses the same protocol as the (native) clients use to access an exclusively owned device. Figure \ref{fig:virtio_qemu} shows the representation of a device virtualization through a multiplexer. In this example each device has a single driver, encapsulated either in a native component or a virtual machine, and is multiplexed securely between clients \footnote{https://trustworthy.systems/projects/TS/drivers/}.
    
    \begin{figure}[hbt]
    \centering
    \includegraphics[scale=0.77]{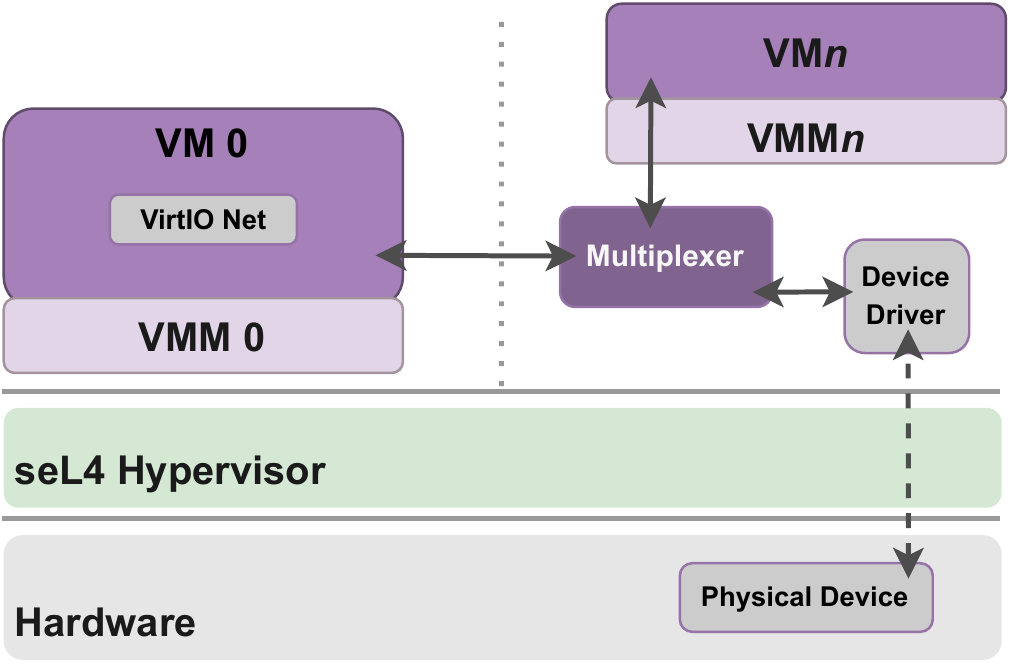}
    \caption{VirtIO interfaces considering a formally verified Device Driver}
    \label{fig:virtio_qemu}
    \end{figure}
    
\end{itemize}

VirtIO also includes standards for Touch, Audio, GPU, and a generic VirtIO-Socket interface which can be used to pass data of any form. Standardized implementations for these are not mature or widely available outside of the automotive use case. OpenSynergy actively worked with Google and Qualcomm to include these interfaces in Android Auto \cite{OPEN22}. It may be possible for the seL4 community to expand those implementations to other areas through customer-funded projects.

%%%%%%%%%%%%%%%%%%%%%%%%%%%%%%%%%%%%%%%%%%
\section{Discussion Topics}
\label{sec:discussion}

\subsection{VMM API}
Apart from the previously mentioned topics, a seL4 standard VMM could also be a programmable API rather than something configured with static Domain Specific Language (DSL) during compilation (e.g., CAmkES). The API makes it possible to wrap the functionality to any compile-time DSLs, custom native services and enables run-time dynamism. The API could have a compile-time configuration for enabling/disabling dynamic features. It should build upon layers so one can use the low-level APIs with all seL4-specific complexity involved, but the API should keep the seL4-specific things minimal at a high level.

An API would make it possible for some elements of the VMM not to be wrapped in a runtime context, like it is now, because it then already makes an assumption about the architecture. That assumption might not be what most integrators (i.e., companies) are after. Let's take KVM as an example. If KVM would provide more than basic constructs and include runtime context (essentially VMM), then we would not be able to have different VMMs (QEMU, \textit{crosvm}, \textit{cloud-hypervisor}). It does not mean that there is not an API already in the seL4 environment. But, it is pretty fragmented and not uniform as one might expect.

The integrators could have an option to use the seL4 VMM (i.e., with characteristics similar to the ones presented in this article) and also the VMM API to have a more diverse virtualization environment. There is a certain minimal subset that a VMM must handle, like handling the hardware virtualization of Generic Interrupt Controller Architecture (GIC) and handling faults. However, it should also be possible to define where VirtIO-console should be handled or that VirtIO-blk device must be handled by QEMU in some VM. If someone has a native VirtIO-backend for some of those examples, it should be possible to use it.

With the seL4 VMM API, it is possible to follow the one VM per VMM "rule" as it is a safer approach from a trust point of view. We could have different flavors of VMMs, such as QEMU, crosvm, and cloud-hypervisor, as each one of them will have its strengths and weakness \cite{RAN20} \cite{ZHA19}. 

\subsection{Formal Methods}
No discussion of an seL4 adjacent system is complete without consideration for the impact of formal methods. Since this discussion is driven by the need for a VMM which can handle complex, real-world use cases, an integrator would likely be using a hardware platform for which seL4 does not yet support formal methods, such as \textit{aarch64} or a multicore configuration. In this case, the effect of formal verification is a moot point. However, in the future, or for a simpler configuration, we can still assess the impact.

Currently, the VMM is assigned per each VM, and thus it is in the VM’s Trusted Computing Base. If we consider the scenario in which it is possible to use a VMM API to run VMMs from different flavors, the formal verification would rely just on the minimal part responsible to execute those VMMs and not in the VMM itself. The VMM is considered part of a guest for the purposes of formal methods, so maintaining the proofs would be challenging. However, there may be a specific case to be made for the standard VMM to be shared across all VMs in a particular system. In that instance, the VMM could be subject to formal methods verification. However, it would be a complex and costly undertaking and goes against the “One VMM Per VM” principle detailed previously in this document. 

Parts of the standard VMM could be subject to verification, an example could be the device drivers. The Device Virtualisation on seL4 project\footnote{https://trustworthy.systems/projects/TS/drivers/devvirt} has the long-term goal of formal verify device drivers, which is enabled by the strong isolation provided for usermode drivers on seL4, which allows verifying drivers in isolation. The seL4 Core Platform has a working in progress project\footnote{https://trustworthy.systems/projects/TS/sel4cp/verification} to formally verify two core aspects of it: (i) correctness of the implementation (i.e. its abstractions function as specified), and (ii) correctness of the system initialisation (i.e. the collection of underlying seL4 objects are fairly represented by the system specification).

%%%%%%%%%%%%%%%%%%%%%%%%%%%%%%%%%%%%%%%%%%
\section{Conclusion}
\label{sec:conc}
Based on the current seL4 VMM implementation, we could conclude that the existing VMM baseline is not ideal and lacks support for many useful design features already present in legacy VMMs of other hypervisors. There are implementations currently in development and supported by the seL4 community, such as the seL4 Core Platform and sDDF. Those approaches may help shrink the gap of current user space seL4 VMM. However, the current gap can be remedied by helping to build a new VMM standard, unified under the principles laid out in this article. Ultimately, after being extended, the present potential directions toward a standard must be put to the test by making a concerted effort to build a real-world proof of concept around it. This will almost certainly require significant funding – either of an R\&D nature or from an end customer. 

Considering the seL4 ecosystem, one step towards defining a standardized VMM would be the creation of an RFC for community discussion and approval. Even though an organizational step, the RFC can start a deeper technical discussion among the seL4 community. It will be up to one or more members of the seL4 community to look for opportunities to take up this mantle and be a champion for this initiative. Also, such standard VMM will only be successful when discussed within the seL4 community. Thus, spreading such ideas through the seL4 community communication channels is essential. In light of this, we started this discussion by presenting a talk and discussion at seL4 Summit 2022\footnote{https://sel4.systems/Foundation/Summit/abstracts2022\#a-Standard-VMM}. From the participation at the seL4 Summit 2022, we could extract potential topics of interest from the seL4 community regarding a standard VMM. Moreover, the creation of work groups within the seL4 Community around topics of interest may be the best approach to leverage such standard VMM.

\conflictsofinterest{The authors declare no conflict of interest.}

%\conflictsofinterest{Declare conflicts of interest or state ``The authors declare no conflict of interest.'' Authors must identify and declare any personal circumstances or interest that may be perceived as inappropriately influencing the representation or interpretation of reported research results. Any role of the funders in the design of the study; in the collection, analyses or interpretation of data; in the writing of the manuscript; or in the decision to publish the results must be declared in this section. If there is no role, please state ``The funders had no role in the design of the study; in the collection, analyses, or interpretation of data; in the writing of the manuscript; or in the decision to publish the~results''.} 

%%%%%%%%%%%%%%%%%%%%%%%%%%%%%%%%%%%%%%%%%%
\begin{adjustwidth}{-\extralength}{0cm}

\reftitle{References}


\bibliography{template}

\begin{thebibliography}{999}

\bibitem[{PR Newswire}(2022)]{BLOOM22}
{PR Newswire}.
\newblock {Embedded System Market Size worth \$159.12 Billion by 2031 CAGR:
  7.7\% - TMR Study},  2022.

\bibitem[Heiser(2008)]{GER08}
Heiser, G.
\newblock The Role of Virtualization in Embedded Systems.
\newblock In Proceedings of the Proceedings of the 1st Workshop on Isolation
  and Integration in Embedded Systems; Association for Computing Machinery: New
  York, NY, USA,  2008; IIES '08, p. 11–16.
\newblock {\url{https://doi.org/10.1145/1435458.1435461}}.

\bibitem[Bauman \em{et~al.}(2015)Bauman, Ayoade, and Lin]{BAU15}
Bauman, E.; Ayoade, G.; Lin, Z.
\newblock A Survey on Hypervisor-Based Monitoring: Approaches, Applications,
  and Evolutions.
\newblock {\em ACM Comput. Surv.} {\bf 2015}, {\em 48}.
\newblock {\url{https://doi.org/10.1145/2775111}}.

\bibitem[Deacon(2019)]{DEA20}
Deacon, W.
\newblock Virtualization for the Masses: Exposing KVM on Android,  2019.

\bibitem[Stabellini(2019)]{STA19}
Stabellini, S.
\newblock True Static Partitioning With Xen Dom0-Less,  2019.

\bibitem[Klein \em{et~al.}(2009)Klein, Elphinstone, Heiser, Andronick, Cock,
  Derrin, Elkaduwe, Engelhardt, Kolanski, Norrish, Sewell, Tuch, and
  Winwood]{HEI09}
Klein, G.; Elphinstone, K.; Heiser, G.; Andronick, J.; Cock, D.; Derrin, P.;
  Elkaduwe, D.; Engelhardt, K.; Kolanski, R.; Norrish, M.;  et~al.
\newblock SeL4: Formal Verification of an OS Kernel.
\newblock In Proceedings of the Proceedings of the ACM SIGOPS 22nd Symposium on
  Operating Systems Principles; Association for Computing Machinery: New York,
  NY, USA,  2009; SOSP '09, p. 207–220.
\newblock {\url{https://doi.org/10.1145/1629575.1629596}}.

\bibitem[Heiser(2020)]{HEI20}
Heiser, G.
\newblock The seL4 Microkernel--An Introduction,  2020.

\bibitem[{seL4 Project}(2022{\natexlab{a}})]{SERV22}
{seL4 Project}.
\newblock Services endorsed by the Foundation,  2022.

\bibitem[{seL4 Project}(2022{\natexlab{b}})]{MEM22}
{seL4 Project}.
\newblock {seL4 Foundation Membership},  2022.

\bibitem[Cinque \em{et~al.}(2022)Cinque, Cotroneo, {De Simone}, and
  Rosiello]{CIN22}
Cinque, M.; Cotroneo, D.; {De Simone}, L.; Rosiello, S.
\newblock Virtualizing mixed-criticality systems: A survey on industrial trends
  and issues.
\newblock {\em Future Generation Computer Systems} {\bf 2022}, {\em
  129},~315--330.
\newblock {\url{https://doi.org/https://doi.org/10.1016/j.future.2021.12.002}}.

\bibitem[Wulf \em{et~al.}(2021)Wulf, Willig, and Göhringer]{WUL21}
Wulf, C.; Willig, M.; Göhringer, D.
\newblock A Survey on Hypervisor-based Virtualization of Embedded
  Reconfigurable Systems.
\newblock In Proceedings of the 2021 31st International Conference on
  Field-Programmable Logic and Applications (FPL),  2021, pp. 249--256.
\newblock {\url{https://doi.org/10.1109/FPL53798.2021.00047}}.

\bibitem[Aalam \em{et~al.}(2021)Aalam, Kumar, and Gour]{AAL21}
Aalam, Z.; Kumar, V.; Gour, S.
\newblock A review paper on hypervisor and virtual machine security.
\newblock {\em Journal of Physics: Conference Series} {\bf 2021}, {\em
  1950},~012027.
\newblock {\url{https://doi.org/10.1088/1742-6596/1950/1/012027}}.

\bibitem[Heiser \em{et~al.}(2020)Heiser, Klein, and Andronick]{GER20}
Heiser, G.; Klein, G.; Andronick, J.
\newblock SeL4 in Australia: From Research to Real-World Trustworthy Systems.
\newblock {\em Commun. ACM} {\bf 2020}, {\em 63},~72–75.
\newblock {\url{https://doi.org/10.1145/3378426}}.

\bibitem[VanderLeest(2018)]{VAND18}
VanderLeest, S.H.
\newblock Is formal proof of seL4 sufficient for avionics security?
\newblock {\em IEEE Aerospace and Electronic Systems Magazine} {\bf 2018}, {\em
  33},~16--21.
\newblock {\url{https://doi.org/10.1109/MAES.2018.160217}}.

\bibitem[VanVossen \em{et~al.}(2019)VanVossen, Millwood, Guikema, Elliott, and
  Roach]{VAN19}
VanVossen, R.; Millwood, J.; Guikema, C.; Elliott, L.; Roach, J.
\newblock The seL4 Microkernel--A Robust, Resilient, and Open-Source Foundation
  for Ground Vehicle Electronics Architecture.
\newblock In Proceedings of the Proceedings of the Ground Vehicle Systems
  Engineering and Technology Symposium,  2019, pp. 13--15.

\bibitem[Millwood \em{et~al.}(2020)Millwood, VanVossen, and Elliott]{MILL20}
Millwood, J.; VanVossen, R.; Elliott, L.
\newblock Performance Impacts from the seL4 Hypervisor.
\newblock In Proceedings of the Proceedings of the Ground Vehicle Systems
  Engineering and Technology Symposium,  2020, pp. 13--15.

\bibitem[Sudvarg and Gill(2022)]{SUD22}
Sudvarg, M.; Gill, C.
\newblock A Concurrency Framework for Priority-Aware Intercomponent Requests in
  CAmkES on seL4.
\newblock In Proceedings of the 2022 IEEE 28th International Conference on
  Embedded and Real-Time Computing Systems and Applications (RTCSA),  2022, pp.
  1--10.
\newblock {\url{https://doi.org/10.1109/RTCSA55878.2022.00007}}.

\bibitem[Elphinstone and Heiser(2013)]{ELP13}
Elphinstone, K.; Heiser, G.
\newblock From L3 to SeL4 What Have We Learnt in 20 Years of L4 Microkernels?
\newblock In Proceedings of the Proceedings of the Twenty-Fourth ACM Symposium
  on Operating Systems Principles; Association for Computing Machinery: New
  York, NY, USA,  2013; SOSP '13, p. 133–150.
\newblock {\url{https://doi.org/10.1145/2517349.2522720}}.

\bibitem[{seL4 Project}(2022{\natexlab{a}})]{FAQ22}
{seL4 Project}.
\newblock {Frequently Asked Questions on seL4},  2022.

\bibitem[{seL4 Project}(2022{\natexlab{b}})]{FOUN22}
{seL4 Project}.
\newblock {seL4 Foundation},  2022.

\bibitem[Roch(2004)]{ROCH04}
Roch, B.
\newblock Monolithic kernel vs. Microkernel.
\newblock {\em TU Wien} {\bf 2004}, {\em 1}.

\bibitem[Rushby(1981)]{RUSH81}
Rushby, J.M.
\newblock Design and Verification of Secure Systems.
\newblock In Proceedings of the Proceedings of the Eighth ACM Symposium on
  Operating Systems Principles; Association for Computing Machinery: New York,
  NY, USA,  1981; SOSP '81, p. 12–21.
\newblock {\url{https://doi.org/10.1145/800216.806586}}.

\bibitem[Nider \em{et~al.}(2019)Nider, Rapoport, and Bottomley]{NID19}
Nider, J.; Rapoport, M.; Bottomley, J.
\newblock Address Space Isolation in the Linux Kernel.
\newblock In Proceedings of the Proceedings of the 12th ACM International
  Conference on Systems and Storage; Association for Computing Machinery: New
  York, NY, USA,  2019; SYSTOR '19, p. 194.
\newblock {\url{https://doi.org/10.1145/3319647.3325855}}.

\bibitem[Biggs \em{et~al.}(2018)Biggs, Lee, and Heiser]{BIG18}
Biggs, S.; Lee, D.; Heiser, G.
\newblock The Jury Is In: Monolithic OS Design Is Flawed: Microkernel-Based
  Designs Improve Security.
\newblock In Proceedings of the Proceedings of the 9th Asia-Pacific Workshop on
  Systems; Association for Computing Machinery: New York, NY, USA,  2018; APSys
  '18.
\newblock {\url{https://doi.org/10.1145/3265723.3265733}}.

\bibitem[Hao \em{et~al.}(2022)Hao, Zhang, Li, Du, Qian, and Sani]{HAO22}
Hao, Y.; Zhang, H.; Li, G.; Du, X.; Qian, Z.; Sani, A.A.
\newblock Demystifying the Dependency Challenge in Kernel Fuzzing.
\newblock In Proceedings of the Proceedings of the 44th International
  Conference on Software Engineering; Association for Computing Machinery: New
  York, NY, USA,  2022; ICSE '22, p. 659–671.
\newblock {\url{https://doi.org/10.1145/3510003.3510126}}.

\bibitem[Hohmuth \em{et~al.}(2004)Hohmuth, Peter, H\"{a}rtig, and
  Shapiro]{HOH04}
Hohmuth, M.; Peter, M.; H\"{a}rtig, H.; Shapiro, J.S.
\newblock Reducing TCB Size by Using Untrusted Components: Small Kernels versus
  Virtual-Machine Monitors.
\newblock In Proceedings of the Proceedings of the 11th Workshop on ACM SIGOPS
  European Workshop; Association for Computing Machinery: New York, NY, USA,
  2004; EW 11, p. 22–es.
\newblock {\url{https://doi.org/10.1145/1133572.1133615}}.

\bibitem[Chiueh and Brook(2005)]{CHI05}
Chiueh, S.N.T.c.; Brook, S.
\newblock A survey on virtualization technologies.
\newblock {\em Rpe Report} {\bf 2005}, {\em 142}.

\bibitem[Popek and Goldberg(1974)]{POP74}
Popek, G.J.; Goldberg, R.P.
\newblock Formal Requirements for Virtualizable Third Generation Architectures.
\newblock {\em Commun. ACM} {\bf 1974}, {\em 17},~412–421.
\newblock {\url{https://doi.org/10.1145/361011.361073}}.

\bibitem[Tiburski \em{et~al.}(2021)Tiburski, Moratelli, Johann, de~Matos, and
  Hessel]{TIB21}
Tiburski, R.T.; Moratelli, C.R.; Johann, S.F.; de~Matos, E.; Hessel, F.
\newblock A lightweight virtualization model to enable edge computing in deeply
  embedded systems.
\newblock {\em Software: Practice and Experience} {\bf 2021}, {\em
  51},~1964--1981,
  \href{http://xxx.lanl.gov/abs/https://onlinelibrary.wiley.com/doi/pdf/10.1002/spe.2968}{{\normalfont
  [https://onlinelibrary.wiley.com/doi/pdf/10.1002/spe.2968]}}.
\newblock {\url{https://doi.org/https://doi.org/10.1002/spe.2968}}.

\bibitem[Moratelli \em{et~al.}(2020)Moratelli, Tiburski, {de Matos}, Portal,
  Johann, and Hessel]{MORA20}
Moratelli, C.R.; Tiburski, R.T.; {de Matos}, E.; Portal, G.; Johann, S.F.;
  Hessel, F.
\newblock Chapter 9 - Privacy and security of Internet of Things devices. In
  {\em Real-Time Data Analytics for Large Scale Sensor Data}; Das, H.; Dey, N.;
  {Emilia Balas}, V., Eds.; Academic Press,  2020; Vol.~6, {\em Advances in
  Ubiquitous Sensing Applications for Healthcare}, pp. 183--214.
\newblock
  {\url{https://doi.org/https://doi.org/10.1016/B978-0-12-818014-3.00009-7}}.

\bibitem[Martins \em{et~al.}(2017)Martins, Alves, Cabral, Tavares, and
  Pinto]{MART17}
Martins, J.; Alves, J.; Cabral, J.; Tavares, A.; Pinto, S.
\newblock μRTZVisor: A Secure and Safe Real-Time Hypervisor.
\newblock {\em Electronics} {\bf 2017}, {\em 6}.
\newblock {\url{https://doi.org/10.3390/electronics6040093}}.

\bibitem[Smith and Nair(2005)]{SMI05}
Smith, J.; Nair, R.
\newblock {\em Virtual machines: versatile platforms for systems and
  processes}; Elsevier,  2005.

\bibitem[Russell(2008)]{RUSS08}
Russell, R.
\newblock Virtio: Towards a de-Facto Standard for Virtual I/O Devices.
\newblock {\em SIGOPS Oper. Syst. Rev.} {\bf 2008}, {\em 42},~95–103.
\newblock {\url{https://doi.org/10.1145/1400097.1400108}}.

\bibitem[Vojnak \em{et~al.}(2019)Vojnak, Ðorđević, Timčenko, and
  Štrbac]{NAK19}
Vojnak, D.T.; Ðorđević, B.S.; Timčenko, V.V.; Štrbac, S.M.
\newblock Performance Comparison of the type-2 hypervisor VirtualBox and VMWare
  Workstation.
\newblock In Proceedings of the 2019 27th Telecommunications Forum (TELFOR),
  2019, pp. 1--4.
\newblock {\url{https://doi.org/10.1109/TELFOR48224.2019.8971213}}.

\bibitem[Azmandian \em{et~al.}(2011)Azmandian, Moffie, Alshawabkeh, Dy, Aslam,
  and Kaeli]{AZM11}
Azmandian, F.; Moffie, M.; Alshawabkeh, M.; Dy, J.; Aslam, J.; Kaeli, D.
\newblock Virtual Machine Monitor-Based Lightweight Intrusion Detection.
\newblock {\em SIGOPS Oper. Syst. Rev.} {\bf 2011}, {\em 45},~38–53.
\newblock {\url{https://doi.org/10.1145/2007183.2007189}}.

\bibitem[Rosenblum and Garfinkel(2005)]{ROS05}
Rosenblum, M.; Garfinkel, T.
\newblock Virtual machine monitors: current technology and future trends.
\newblock {\em Computer} {\bf 2005}, {\em 38},~39--47.
\newblock {\url{https://doi.org/10.1109/MC.2005.176}}.

\bibitem[Tickoo \em{et~al.}(2010)Tickoo, Iyer, Illikkal, and Newell]{TIC10}
Tickoo, O.; Iyer, R.; Illikkal, R.; Newell, D.
\newblock Modeling Virtual Machine Performance: Challenges and Approaches.
\newblock {\em SIGMETRICS Perform. Eval. Rev.} {\bf 2010}, {\em 37},~55–60.
\newblock {\url{https://doi.org/10.1145/1710115.1710126}}.

\bibitem[Xu \em{et~al.}(2014)Xu, Liu, Jin, and Vasilakos]{XU14}
Xu, F.; Liu, F.; Jin, H.; Vasilakos, A.V.
\newblock Managing Performance Overhead of Virtual Machines in Cloud Computing:
  A Survey, State of the Art, and Future Directions.
\newblock {\em Proceedings of the IEEE} {\bf 2014}, {\em 102},~11--31.
\newblock {\url{https://doi.org/10.1109/JPROC.2013.2287711}}.

\bibitem[Barham \em{et~al.}(2003)Barham, Dragovic, Fraser, Hand, Harris, Ho,
  Neugebauer, Pratt, and Warfield]{BAR03}
Barham, P.; Dragovic, B.; Fraser, K.; Hand, S.; Harris, T.; Ho, A.; Neugebauer,
  R.; Pratt, I.; Warfield, A.
\newblock Xen and the Art of Virtualization.
\newblock In Proceedings of the Proceedings of the Nineteenth ACM Symposium on
  Operating Systems Principles; Association for Computing Machinery: New York,
  NY, USA,  2003; SOSP '03, p. 164–177.
\newblock {\url{https://doi.org/10.1145/945445.945462}}.

\bibitem[Chierici and Veraldi(2010)]{CHI10}
Chierici, A.; Veraldi, R.
\newblock A quantitative comparison between xen and kvm.
\newblock {\em Journal of Physics: Conference Series} {\bf 2010}, {\em
  219},~042005.
\newblock {\url{https://doi.org/10.1088/1742-6596/219/4/042005}}.

\bibitem[Dall and Nieh(2014)]{DALL14}
Dall, C.; Nieh, J.
\newblock KVM/ARM: The Design and Implementation of the Linux ARM Hypervisor.
\newblock {\em SIGPLAN Not.} {\bf 2014}, {\em 49},~333–348.
\newblock {\url{https://doi.org/10.1145/2644865.2541946}}.

\bibitem[Raho \em{et~al.}(2015)Raho, Spyridakis, Paolino, and Raho]{RAH15}
Raho, M.; Spyridakis, A.; Paolino, M.; Raho, D.
\newblock KVM, Xen and Docker: A performance analysis for ARM based NFV and
  cloud computing.
\newblock In Proceedings of the 2015 IEEE 3rd Workshop on Advances in
  Information, Electronic and Electrical Engineering (AIEEE),  2015, pp. 1--8.
\newblock {\url{https://doi.org/10.1109/AIEEE.2015.7367280}}.

\bibitem[Mansouri and Babar(2021)]{YAS21}
Mansouri, Y.; Babar, M.A.
\newblock A review of edge computing: Features and resource virtualization.
\newblock {\em Journal of Parallel and Distributed Computing} {\bf 2021}, {\em
  150},~155--183.
\newblock {\url{https://doi.org/https://doi.org/10.1016/j.jpdc.2020.12.015}}.

\bibitem[Ramalho and Neto(2016)]{RAM16}
Ramalho, F.; Neto, A.
\newblock Virtualization at the network edge: A performance comparison.
\newblock In Proceedings of the 2016 IEEE 17th International Symposium on A
  World of Wireless, Mobile and Multimedia Networks (WoWMoM),  2016, pp. 1--6.
\newblock {\url{https://doi.org/10.1109/WoWMoM.2016.7523584}}.

\bibitem[Hwang \em{et~al.}(2008)Hwang, Suh, Heo, Park, Ryu, Park, and
  Kim]{HW08}
Hwang, J.Y.; Suh, S.B.; Heo, S.K.; Park, C.J.; Ryu, J.M.; Park, S.Y.; Kim, C.R.
\newblock Xen on ARM: System Virtualization Using Xen Hypervisor for ARM-Based
  Secure Mobile Phones.
\newblock In Proceedings of the 2008 5th IEEE Consumer Communications and
  Networking Conference,  2008, pp. 257--261.
\newblock {\url{https://doi.org/10.1109/ccnc08.2007.64}}.

\bibitem[Stabellini and Campbell(2012)]{STA12}
Stabellini, S.; Campbell, I.
\newblock Xen on arm cortex a15.
\newblock {\em Xen Summit North America} {\bf 2012}, {\em 2012}.

\bibitem[Martins \em{et~al.}(2020)Martins, Tavares, Solieri, Bertogna, and
  Pinto]{MAR20}
Martins, J.; Tavares, A.; Solieri, M.; Bertogna, M.; Pinto, S.
\newblock {Bao: A Lightweight Static Partitioning Hypervisor for Modern
  Multi-Core Embedded Systems}.
\newblock In Proceedings of the Workshop on Next Generation Real-Time Embedded
  Systems (NG-RES 2020); Bertogna, M.; Terraneo, F., Eds.; Schloss
  Dagstuhl--Leibniz-Zentrum fuer Informatik: Dagstuhl, Germany,  2020; Vol.~77,
  {\em OpenAccess Series in Informatics (OASIcs)}, pp. 3:1--3:14.
\newblock {\url{https://doi.org/10.4230/OASIcs.NG-RES.2020.3}}.

\bibitem[Li \em{et~al.}(2019)Li, Xu, Ren, and Dong]{LI19}
Li, H.; Xu, X.; Ren, J.; Dong, Y.
\newblock ACRN: A Big Little Hypervisor for IoT Development.
\newblock In Proceedings of the Proceedings of the 15th ACM SIGPLAN/SIGOPS
  International Conference on Virtual Execution Environments; Association for
  Computing Machinery: New York, NY, USA,  2019; VEE 2019, p. 31–44.
\newblock {\url{https://doi.org/10.1145/3313808.3313816}}.

\bibitem[Li \em{et~al.}(2021)Li, Li, Gu, Nieh, and Hui]{WEI21}
Li, S.W.; Li, X.; Gu, R.; Nieh, J.; Hui, J.Z.
\newblock Formally Verified Memory Protection for a Commodity Multiprocessor
  Hypervisor.
\newblock In Proceedings of the 30th USENIX Security Symposium (USENIX Security
  21). USENIX Association,  2021, pp. 3953--3970.

\bibitem[Bellard(2005)]{BELL05}
Bellard, F.
\newblock {QEMU}, a Fast and Portable Dynamic Translator.
\newblock In Proceedings of the 2005 USENIX Annual Technical Conference (USENIX
  ATC 05); USENIX Association: Anaheim, CA,  2005.

\bibitem[Agache \em{et~al.}(2020)Agache, Brooker, Iordache, Liguori,
  Neugebauer, Piwonka, and Popa]{AGA20}
Agache, A.; Brooker, M.; Iordache, A.; Liguori, A.; Neugebauer, R.; Piwonka,
  P.; Popa, D.M.
\newblock Firecracker: Lightweight Virtualization for Serverless Applications.
\newblock In Proceedings of the 17th USENIX Symposium on Networked Systems
  Design and Implementation (NSDI 20); USENIX Association: Santa Clara, CA,
  2020; pp. 419--434.

\bibitem[Tsirkin and Huck(2018)]{VIRT18}
Tsirkin, M.S.; Huck, C.
\newblock Virtual I/O Device (VIRTIO) Version 1.1.
\newblock {\em OASIS Committee} {\bf 2018}.

\bibitem[{seL4 Project}(2022{\natexlab{a}})]{VIR22}
{seL4 Project}.
\newblock {Virtualisation on seL4},  2022.

\bibitem[{seL4 Project}(2022{\natexlab{b}})]{CAMVMM22}
{seL4 Project}.
\newblock {CAmkES VMM},  2022.

\bibitem[{seL4 Project}(2022{\natexlab{c}})]{CAM22}
{seL4 Project}.
\newblock {CAmkES},  2022.

\bibitem[seL4 Project(2022)]{COR22}
seL4 Project.
\newblock {The seL4 Core Platform},  2022.

\bibitem[Leslie and Heiser(2022{\natexlab{a}})]{COR22b}
Leslie, B.; Heiser, G.
\newblock {The seL4 Core Platform},  2022.

\bibitem[Leslie and Heiser(2022{\natexlab{b}})]{COR22c}
Leslie, B.; Heiser, G.
\newblock {Evolving seL4CP Into a Dynamic OS},  2022.

\bibitem[Sherman(2008)]{SHER08}
Sherman, T.
\newblock Quality Attributes for Embedded Systems.
\newblock In Proceedings of the Advances in Computer and Information Sciences
  and Engineering; Sobh, T., Ed.; Springer Netherlands: Dordrecht,  2008; pp.
  536--539.

\bibitem[Clements \em{et~al.}(2003)Clements, Kazman, Klein, et~al.]{CLE03}
Clements, P.; Kazman, R.; Klein, M.;  et~al.
\newblock {\em Evaluating software architectures}; Tsinghua University Press
  Beijing,  2003.

\bibitem[Oliveira \em{et~al.}(2013)Oliveira, Guessi, Feitosa, Manteuffel,
  Galster, Oquendo, and Nakagawa]{OLI13}
Oliveira, L.; Guessi, M.; Feitosa, D.; Manteuffel, C.; Galster, M.; Oquendo,
  F.; Nakagawa, E.
\newblock An Investigation on Quality Models and Quality Attributes for
  Embedded Systems.
\newblock In Proceedings of the The Eight International Conference on Software
  Engineering Advances, ICSEA ; Conference date: 27-10-2013 Through 31-10-2013,
   2013, pp. 523--528.

\bibitem[Bianchi \em{et~al.}(2015)Bianchi, Santos, and Felizardo]{BIA15}
Bianchi, T.; Santos, D.S.; Felizardo, K.R.
\newblock Quality Attributes of Systems-of-Systems: A Systematic Literature
  Review.
\newblock In Proceedings of the 2015 IEEE/ACM 3rd International Workshop on
  Software Engineering for Systems-of-Systems,  2015, pp. 23--30.
\newblock {\url{https://doi.org/10.1109/SESoS.2015.12}}.

\bibitem[Li \em{et~al.}(2022)Li, Li, Dall, Gu, Nieh, Sait, and
  Stockwell]{XUP22}
Li, X.; Li, X.; Dall, C.; Gu, R.; Nieh, J.; Sait, Y.; Stockwell, G.
\newblock Design and Verification of the Arm Confidential Compute Architecture.
\newblock In Proceedings of the 16th USENIX Symposium on Operating Systems
  Design and Implementation (OSDI 22); USENIX Association: Carlsbad, CA,  2022;
  pp. 465--484.

\bibitem[Vázquez-Ingelmo \em{et~al.}(2020)Vázquez-Ingelmo, García-Holgado,
  and García-Peñalvo]{VAZ20}
Vázquez-Ingelmo, A.; García-Holgado, A.; García-Peñalvo, F.J.
\newblock C4 model in a Software Engineering subject to ease the comprehension
  of UML and the software.
\newblock In Proceedings of the 2020 IEEE Global Engineering Education
  Conference (EDUCON),  2020, pp. 919--924.
\newblock {\url{https://doi.org/10.1109/EDUCON45650.2020.9125335}}.

\bibitem[seL4 Project(2022)]{SUP22}
seL4 Project.
\newblock Supported Platforms,  2022.

\bibitem[Pinto and Santos(2019)]{PIN19}
Pinto, S.; Santos, N.
\newblock Demystifying Arm TrustZone: A Comprehensive Survey.
\newblock {\em ACM Comput. Surv.} {\bf 2019}, {\em 51}.
\newblock {\url{https://doi.org/10.1145/3291047}}.

\bibitem[Kuz \em{et~al.}(2007)Kuz, Liu, Gorton, and Heiser]{KUZ07}
Kuz, I.; Liu, Y.; Gorton, I.; Heiser, G.
\newblock CAmkES: A component model for secure microkernel-based embedded
  systems.
\newblock {\em Journal of Systems and Software} {\bf 2007}, {\em 80},~687--699.
\newblock Component-Based Software Engineering of Trustworthy Embedded Systems,
  {\url{https://doi.org/https://doi.org/10.1016/j.jss.2006.08.039}}.

\bibitem[{Open Synergy}(2022)]{OPEN22}
{Open Synergy}.
\newblock Android Ecosystem,  2022.

\bibitem[Randal(2020)]{RAN20}
Randal, A.
\newblock The Ideal Versus the Real: Revisiting the History of Virtual Machines
  and Containers.
\newblock {\em ACM Comput. Surv.} {\bf 2020}, {\em 53}.
\newblock {\url{https://doi.org/10.1145/3365199}}.

\bibitem[Zhang \em{et~al.}(2019)Zhang, Zheng, Wang, Li, Fu, Zhang, and
  Shen]{ZHA19}
Zhang, X.; Zheng, X.; Wang, Z.; Li, Q.; Fu, J.; Zhang, Y.; Shen, Y.
\newblock Fast and Scalable VMM Live Upgrade in Large Cloud Infrastructure.
\newblock In Proceedings of the Proceedings of the Twenty-Fourth International
  Conference on Architectural Support for Programming Languages and Operating
  Systems; Association for Computing Machinery: New York, NY, USA,  2019;
  ASPLOS '19, p. 93–105.
\newblock {\url{https://doi.org/10.1145/3297858.3304034}}.

\end{thebibliography}

\end{adjustwidth}
\end{document}